\documentclass[acmtog, authorversion]{acmart}

\usepackage{booktabs} 

\usepackage{CJKutf8}
\usepackage{subcaption}

\setcitestyle{square}

\usepackage[ruled]{algorithm2e} 

\SetAlFnt{\small}
\SetAlCapFnt{\small}
\SetAlCapNameFnt{\small}
\SetAlCapHSkip{0pt}

\acmJournal{TOG}

\begin{document}

\begin{CJK}{UTF8}{}
 \CJKfamily{mj}
 
\title{Efficient Cloth Simulation using Miniature Cloth and Upscaling Deep Neural Networks}

\author{Tae Min Lee}
\author{Young Jin Oh}
\author{In-Kwon Lee}
\affiliation{%
  \institution{Yonsei University}
  \department{Computer Science}
  \city{Seoul}
}
\email{{dnflxoals, skrcjstk}@gmail.com}
\email{iklee@yonsei.ac.kr}

\renewcommand\shortauthors{Lee, T. et al.}

\begin{abstract}

Cloth simulation requires a fast and stable method for interactively and realistically visualizing fabric materials using computer graphics. We propose an efficient cloth simulation method using miniature cloth simulation and upscaling Deep Neural Networks (DNN). The upscaling DNNs generate the target cloth simulation from the results of physically-based simulations of a miniature cloth that has similar physical properties to those of the target cloth. We have verified the utility of the proposed method through experiments, and the results demonstrate that it is possible to generate fast and stable cloth simulations under various conditions.

\end{abstract}

%
%


%
%


\maketitle

\section{Introduction}

Cloth simulation is an essential technology that has been used to model various fabric materials in the field of computer graphics, such as movies and games. Therefore, many methods for simulating cloth naturally and realistically have been studied~\cite{Terzopoulos:1987:EDM:37401.37427, Baraff:1998:LSC:280814.280821, Choi:2005:SBR:1198555.1198571, Kaldor:2008:SKC:1360612.1360664, Weidner:2018:ECS:3197517.3201281}. In addition, because real-time interactive graphics content is required in research areas such as augmented and virtual reality applications, studies of efficient cloth simulation methods have been continuously carried out~\cite{vassilev2001fast, oh2008physically, BENDER2013945}.

\begin{figure*}[tbp]
\centering
\begin{subfigure}{0.48\textwidth}
\includegraphics[width=\columnwidth]{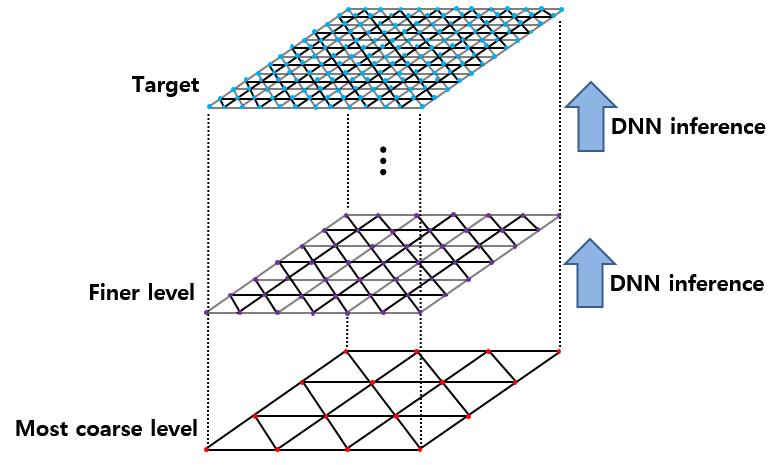}
\caption{Hierarchical cloth model}
\end{subfigure}
\begin{subfigure}{0.48\textwidth}
\includegraphics[width=\columnwidth]{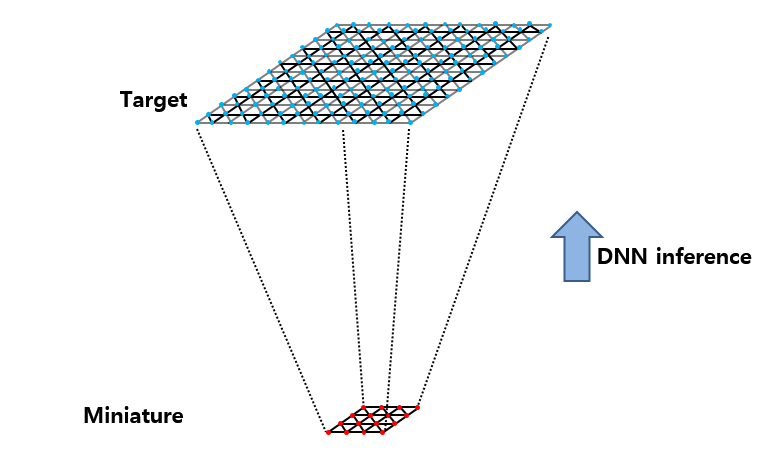}
\caption{Miniature cloth model}
\end{subfigure}
\caption{Structure difference between hierarchical cloth in H-DNN~\cite{Oh:2018:HCS:3208159.3208162} and our miniature cloth model: (a) hierarchical cloth model and (b) miniature cloth model.}
\label{fig:cloth_model}
\end{figure*}


Recently, in the field of physically-based simulations aside from cloth simulation, studies using Deep Neural Networks (DNN) have been actively carried out in an effort to reduce the long computation times or improve the simulation quality.
For example, the DNN inference is used to replace some of the complex computations in the physically-based simulation with the inference process of a neural network~\cite{yang2016data, tompson2016accelerating} or to generate fine-grained simulations based on coarse-grained simulations~\cite{Chu:2017:DSS:3072959.3073643, Xie:2018:TTC:3197517.3201304}. 
However, most studies using DNN for physically-based simulation have focused on fluid simulation. Because it is difficult to apply previous methods directly to cloth simulation, further investigation of the DNN model for cloth simulation is necessary.


To perform efficient cloth simulation using DNN, we must use a network model with low computational cost. We could use Convolutional Neural Networks (CNN), which are frequently used with high accuracy and quality in the field of computer vision \cite{NIPS2012_4824,simonyan2014very,radford2015unsupervised,8237506}, for cloth simulation. However, because these neural network models have high computational costs arising from the large number of iterative convolution calculations, it is challenging to simulate cloth in a more efficient manner than that of conventional physically-based simulation.


Even if the cloth simulation is computed efficiently, the result should be of a similar quality to that generated from existing physically-based simulation methods. The recently proposed H-DNN (hierarchical cloth simulation using DNN) method~\cite{Oh:2018:HCS:3208159.3208162} constructs a hierarchy of cloth particles and performs a physical simulation at the coarsest level. It is then possible to compute a finer-level cloth simulation through DNN inference (see Figure~\ref{fig:cloth_model}(a)). Although this method results in efficient computation times, the quality of the simulations is much lower than that of existing physically-based simulations, mainly because the physical characteristics of the coarsest cloth model (e.g., the rest length between the cloth particles) are not the same as those of the close in the finest model. \cite{Feng:2010:DTR:1778765.1778845,Kavan:2011:PUC:2010324.1964988}.

%


The cloth simulation method proposed in this paper efficiently generates results using a miniature cloth and upscaling DNN.
The miniature cloth is a down-sampled and down-scaled version of the target cloth that we want to simulate. The miniature cloth is simulated by the conventional physically-based simulation method. Based on the results of the miniature cloth simulation, the upscaling DNN yields the target cloth simulation (see Figure~\ref{fig:cloth_model}(b)). This is similar to cinematographic methods, where miniature environments are created for filming by reducing the target-scale objects when it is difficult or expensive to film a full-scale environment~\cite{fielding2013techniques}. Just as the results from filming a miniature environment are expressed in the target-scale via post-processing, we use upscaling DNN inference to generate a simulation of the target cloth based on the simulation of the miniature cloth.


To obtain results from the proposed method that are similar to those from the conventional physically-based simulation, we use a miniature cloth that maintains the same rest length as the target cloth. Because the miniature cloth simulation shows similar movements to the target cloth simulation, high-frequency details that are similar to the target cloth simulation are sufficiently preserved.
In addition, we can simulate a miniature cloth using fewer particles than in the target cloth, and the upscaling DNN can be a lightweight neural network with low computational cost, allowing for more efficient simulation than conventional methods.


This paper is organized as follows. Section~\ref{sec:related-work} discusses works related to cloth simulation and physically-based simulation with DNN. In Section~\ref{sec:upscalingDNN}, we present the training method and architecture of the upscaling DNN. In Section~\ref{sec:proposed-method}, we present the process for target cloth simulation. In Section~\ref{sec:results}, we discuss the performance of the proposed method based on the obtained results. Finally, Section~\ref{sec:conclusion} presents our conclusions and discusses possible future work.

\section{Related Work}
\label{sec:related-work}

\subsection{Cloth Simulation}

Physically-based simulation methods for realistically expressing the movement of elastic materials have also been extensively studied for cloth simulation~\cite{Terzopoulos:1987:EDM:37401.37427, choi2005research}. Breen et al.~\cite{Breen:1994:PDW:192161.192259} proposed a mass-spring system for cloth simulation, and Provoet et al.~\cite{provot1995deformation} proposed a cloth simulation method using explicit Euler integration. Cloth simulation using explicit Euler integration is able to generate real-time interactive results because rapid calculation is possible, although this method did not produce stable results when simulated with a large time step. Therefore, a method using implicit Euler integration was proposed that generates stable results even when simulated with a large time step~\cite{Baraff:1998:LSC:280814.280821}. To accelerate the implicit Euler integration, which has a high computational cost, Liu et al.~\cite{Liu:2013:FSM:2508363.2508406} proposed a method using a solver based on the block coordinate descent scheme. Projective dynamics proposed by Bouaziz et al.~\cite{Bouaziz:2014:PDF:2601097.2601116} extended the previous techniques, which were limited to the mass-spring system, allowing their application to various constrained dynamics problems. To solve the optimization problems in projective dynamics quickly, a method using the quasi-Newton optimization algorithm~\cite{Liu:2017:QMR:3087678.2990496} and another method applying Anderson acceleration~\cite{Peng:2018:AAG:3197517.3201290} have been proposed. Overby et al.~\cite{7990052} applied the Alternating Direction Method of Multipliers (ADMM) to solve the optimization problems in projective dynamics quickly and generate stable results, even when various constraints were changed dynamically.

\subsection{Physically-based Simulation with Deep Neural Networks}

In recent years, studies applying DNN, which has excellent performance in approximating complex computation processes, to physically-based simulation have been actively carried out. Among them, methods using DNN to train the physical movements of simple rigid or elastic bodies and to predict future movements directly have been proposed. A neural physics engine~\cite{chang2016compositional} was introduced to train intuitive physics based on a real scene and the past states of a rigid body, then generating a physical motion sequence. Visual interaction networks~\cite{NIPS2017_7040} were proposed to predict the next positions of an elastic body based on a training process using a physically-based simulation video. These studies are limited to simple structures and predict future movements, but it is difficult to apply these methods to predict the changes in rigid or elastic bodies with complex structures.


Unlike methods in which DNN directly predicts future movements, methods using DNN to facilitate faster computation for existing physically-based simulation methods have also been proposed. Tompson et al.~\cite{tompson2016accelerating} proposed a method for replacing the pressure projection step, which has the highest computational cost, with DNN inference for fast fluid simulation calculations. Wiewel et al.~\cite{wiewel2018latent} proposed a long short-term memory-based method for predicting changes in fluid pressure fields. Luo et al.~\cite{luo2018deepwarp} suggested a method for warping linear elastic simulations into nonlinear elastic simulations via a DNN inference.


Additionally, methods for using DNN inference combined with low-resolution results from physically-based simulations have been proposed to reduce the computational cost of generating high-resolution simulations. In the field of fluid simulation, Chu and Thuerey~\cite{Chu:2017:DSS:3072959.3073643} proposed a CNN-based method to obtain low-resolution simulations for the generation of high-resolution results. Xie et al.~\cite{Xie:2018:TTC:3197517.3201304} proposed a technique for converting low-resolution grid simulations into high-resolution results by adding a temporal coherence discriminator to a super-resolution generative adversarial networks~\cite{8099502}. In the field of cloth simulation, Oh et al.~\cite{Oh:2018:HCS:3208159.3208162} proposed a method for calculating the positions of particles at the coarsest-level using physically-based simulation and then predicting the positions of finer-level particles via DNN inference. This method (we call this method H-DNN) uses the coarsest-level simulation with a small number of particles and a lightweight neural network for greater simulation efficiency compared to that of conventional physically-based simulation methods. However, because the physical properties of the coarsest-level and target cloth are different, even if a finer-level is predicted, the results show that the high-frequency details are not sufficiently preserved.

\begin{figure}[tbp]
  \centering
  \includegraphics[width=\columnwidth]{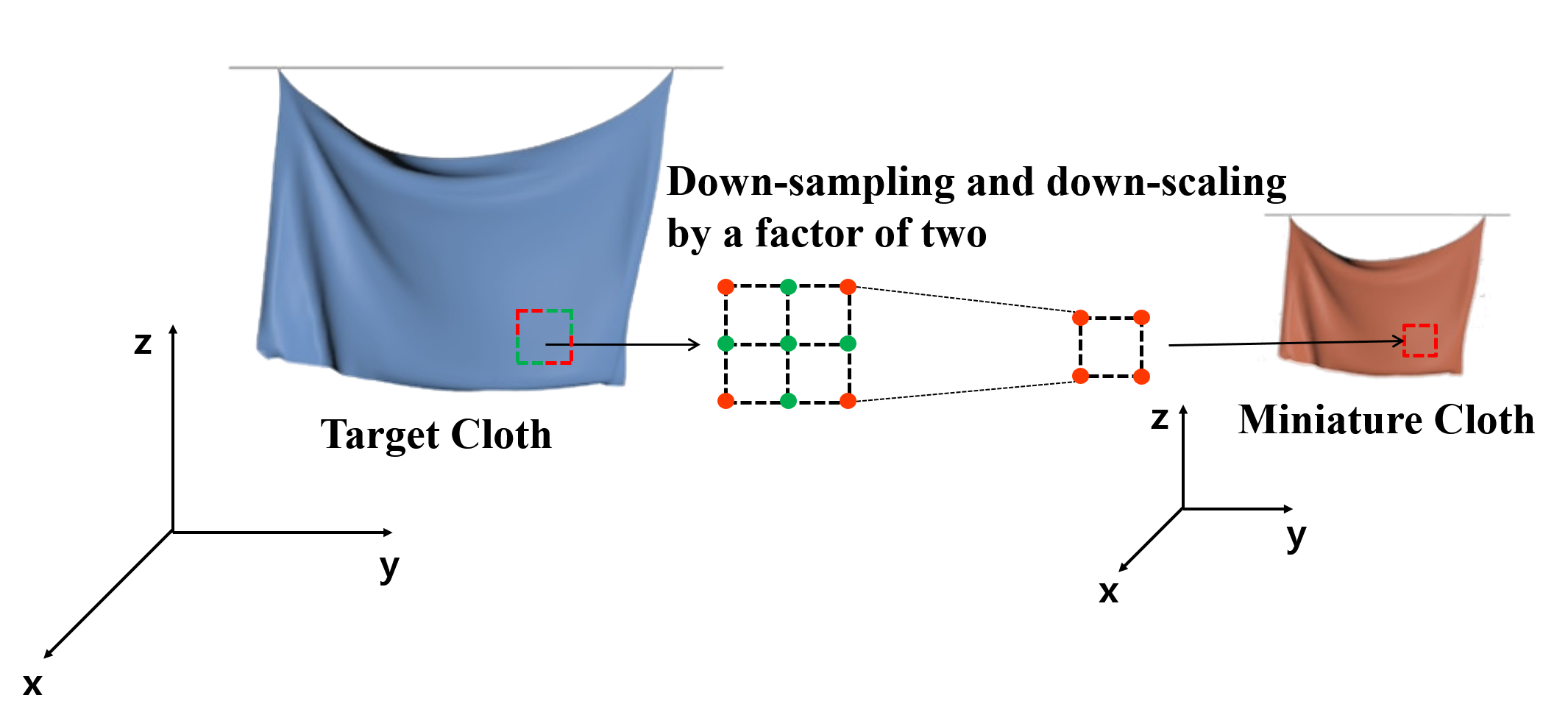}
  \caption{Process of creating a miniature cloth system.}
  \label{fig:downsampling}
\end{figure}

\begin{figure*}
  \centering
  \includegraphics[width=\linewidth]{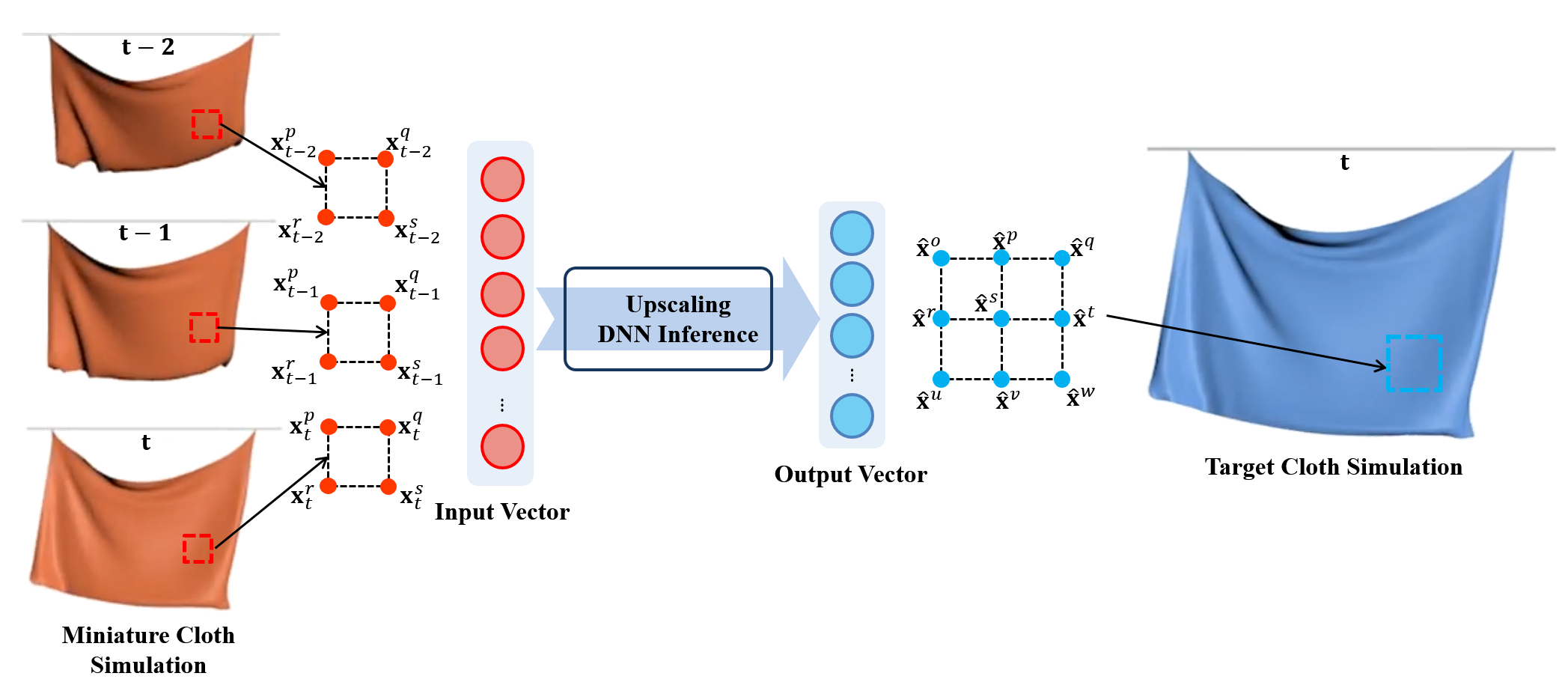}
  \caption{Input and output feature vectors of the upscaling DNN.}
  \label{fig:inference}
\end{figure*}

\section{Upscaling Deep Neural Networks}
\label{sec:upscalingDNN}


The upscaling DNN generates the target cloth simulation based on the simulation of a miniature cloth that is obtained via the down-sampling and down-scaling (DSDS) of the target cloth. To obtain training data for the upscaling DNN, target cloth simulations are carried out via conventional physically-based simulation method, and the results of miniature cloth simulation are generated via down-sampling and down-scaling. We designed and used lightweight neural networks that have a low computational cost to simulate the target cloth efficiently.

\subsection{Miniature Cloth}

Figure~\ref{fig:downsampling} shows the process of creating a miniature cloth system, which is obtained by applying a down-sampling and down-scaling (DSDS) step with a factor of two to the target cloth system. In the down-sampling process, a $ 3 \times 3 $ patch consisting of nine particles from the target cloth is reduced to a $ 2 \times 2 $ patch consisting of four particles in the miniature cloth. In the down-scaling process, the axes of the miniature cloth system are defined as the axes of the target cloth system down-scaled by a factor of two, and the miniature cloth simulation takes place in this down-scaled environment. Thus, the distance between particles, which is one of the most critical factors determining the implicit forces of the cloth system, is maintained at the same value as that in the target cloth system. This is because the miniature cloth is intended to have similar physical properties to those of the target cloth, meaning that the miniature cloth simulation shows similar movements to the target cloth simulation. 
Table~\ref{tab:particles} shows the miniature cloth creation results of three examples that we used to verify our method in this study. The number of particles in the miniature cloth in each example is reduced by approximately 1/4 of that in the target cloth when we apply a DSDS factor of 2.

\begin{table}[t]
\centering
\caption{The number of particles in the miniature cloth and target cloth in each example. The number of particles in the miniature cloth is approximately $1/4$ of that in the target cloth.}
\label{tab:particles}
\renewcommand{\arraystretch}{1.2}
\begin{tabular}{ccc}
\hline
Example   & \#Particles of Miniature & \#Particles of Target \\
\hline
Curtain   & 925              & 3577          \\
Flag      & 425              & 1617          \\
Collision & 651              & 2501          \\         
\hline
\end{tabular}
\end{table}



\begin{figure}[tbp]
  \centering
  \includegraphics[width=\columnwidth]{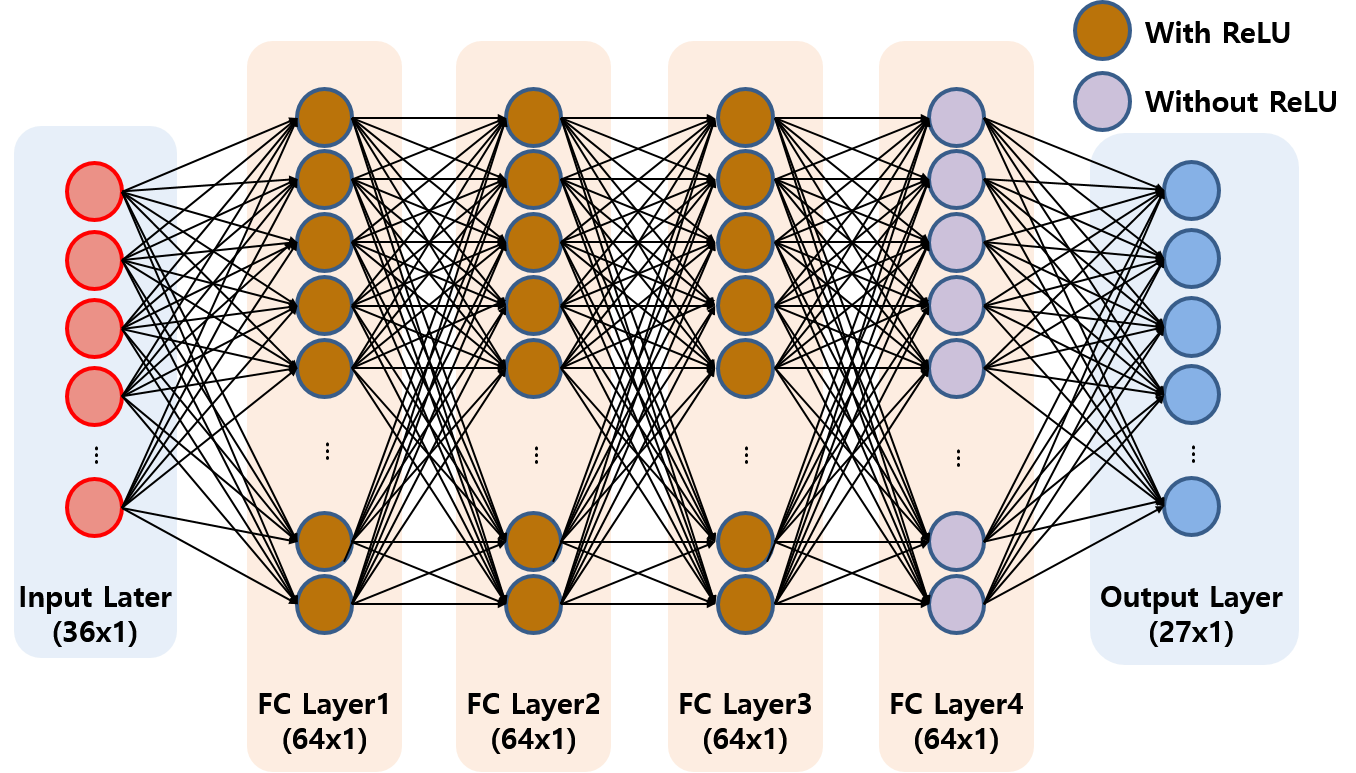}
  \caption{Upscaling Deep Neural Networks model.}
  \label{fig:DNN}
\end{figure}

\begin{figure*}[htbp]
  \centering
  \includegraphics[width=0.9\linewidth]{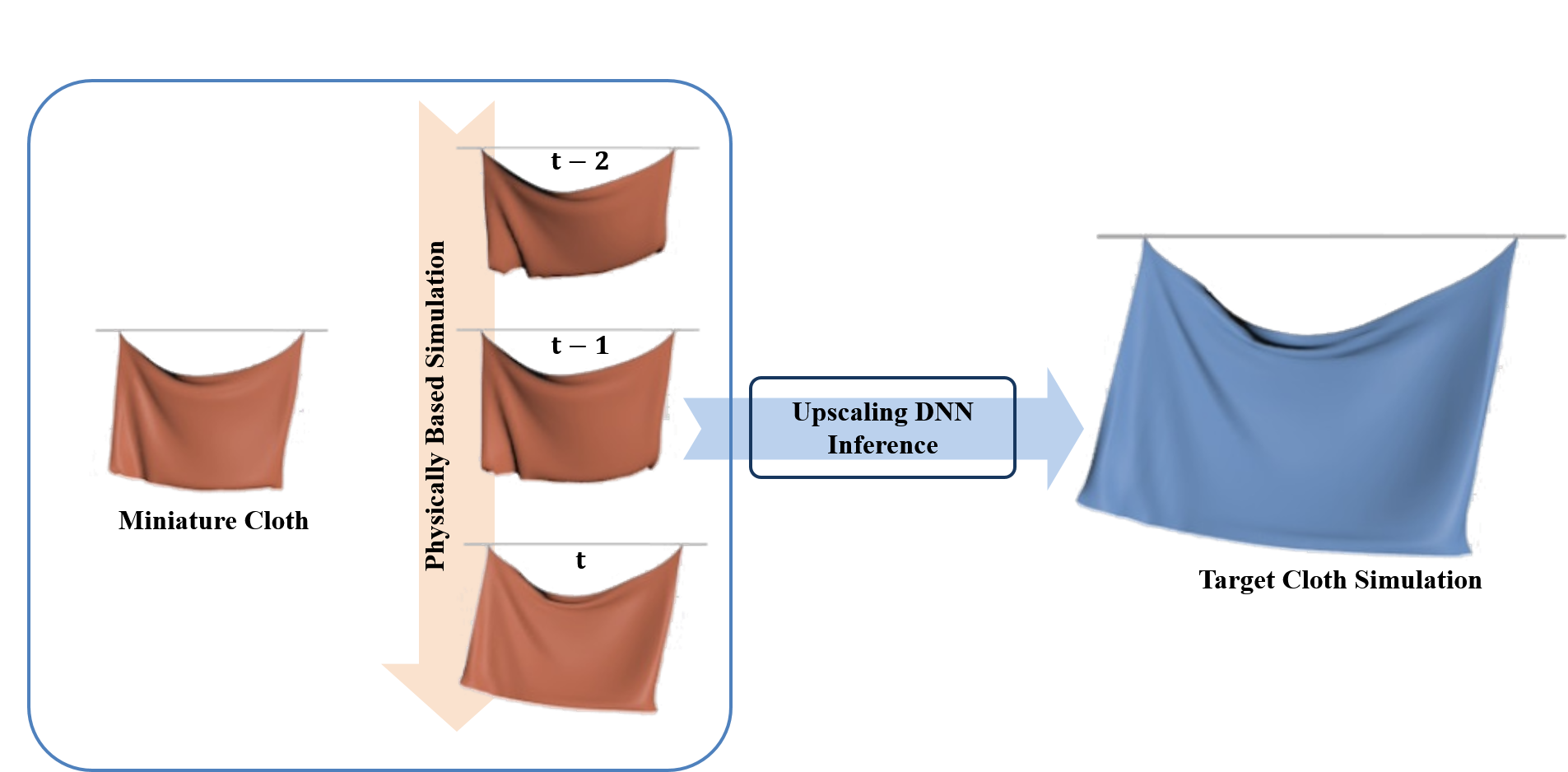}
  \caption{Target cloth simulation generation.}
  \label{fig:overview}
\end{figure*}

\subsection{Input and Output Feature Vectors of Upscaling DNN}

The input and output feature vectors of the upscaling DNN are defined as the positions of the particles in the patches of the miniature and target cloths, respectively. H-DNN~\cite{Oh:2018:HCS:3208159.3208162} uses the positions of the particles in a single frame as feature values. This makes it difficult for DNN to preserve the time coherence between consecutive frames naturally. In contrast, our method uses the concatenated positions of three consecutive frames, including the current and the previous two frames, as input feature vectors. This allows DNN to generate more natural results by taking into account the smooth transition between the neighboring frames. 
In addition, using the particle positions of consecutive frames in an input vector has the effect of reflecting the moving particles' velocity on the input. We have tested several combinations of various features, for example, including the velocity of particles as features. Finally, we found that this simple method of combining the particle positions of several frames was the most effective (see Section~\ref{subsec:performance-features}).


Figure~\ref{fig:inference} illustrates the input and output feature vectors of the proposed DNN model. 
Consider the positions of four particles $\mathbf{x}^p$, $\mathbf{x}^q$, $\mathbf{x}^r$, and $\mathbf{x}^s$ in a $2 \times 2$ particle patch in a miniature cloth. The state vector of the patch $\mathbf{s}$ is defined as follows:
\begin{equation}
\mathbf{s}=[\mathbf{x}^p, \mathbf{x}^q, \mathbf{x}^r, \mathbf{x}^s].
\end{equation}
The input feature vector $\pmb{\delta}_t$ at the simulation time $t$ is defined as
\begin{equation}
\pmb{\delta}_t=\left [\mathbf{s}_{t-2}, \mathbf{s}_{t-1}, \mathbf{s}_t \right ],
\end{equation}
which concatenates the state vectors of three consecutive simulation results at time $t-2$, $t-1$, and $t$.
The output feature vector $\mathbf{o}_t$ at the simulation time $t$ is defined as
\begin{equation}
\mathbf{o}_t=\left [\mathbf{\hat x}^o, \mathbf{\hat x}^p, \mathbf{\hat x}^q, \mathbf{\hat x}^r, \mathbf{\hat x}^s, \mathbf{\hat x}^t, \mathbf{\hat x}^u, \mathbf{\hat x}^v, \mathbf{\hat x}^w \right ],
\end{equation}
which represents the positions of the nine particles in a $3 \times 3$ particle patch in the target cloth that corresponds to a $2 \times 2$ particle patch in the miniature cloth.

\begin{table}[t]
\centering
\caption{The number of input feature and ground-truth output pairs collected in each example and the corresponding number of frames.}
\label{tab:training_data}
\renewcommand{\arraystretch}{1.2}
\begin{tabular}{ccc}
\hline
Example   & \#Input feature and ground-truth pairs & \#Frames \\
\hline
Curtain   & 4,211,294              & 1,262          \\
Flag      & 4,051,917              & 2,781          \\
Collision & 4,652,622              & 2,022          \\         
\hline
\end{tabular}
\end{table}

\subsection{Network Model}
\label{subsec:DNN_structure}

Figure~\ref{fig:DNN} shows the structure of the upscaling DNN model used in the proposed method. We designed a lightweight network consisting of four Fully Connected (FC) layers with a Rectified Linear Unit (ReLU) function as the activation function for fast inference. The first three FC layers calculate intermediate vectors of size 64, and the final FC layer calculates the output feature vectors. The size of the input feature vectors is 36 because it represents the $x, y, z$ position of the three frames of the particle patch in the miniature cloth. The size of the output feature vectors is 27 because it represents the $x, y, z$ position for one frame of the corresponding particle patch in the target cloth.


We define the loss function as the Mean Squared Error (MSE) of the corresponding ground-truth output vector $\mathbf{g}$ and output vector $\mathbf{o}$ computed by inference as follows:
\begin{equation}
Loss = \frac{1}{N}\sum_{n=1}^{N}\sum_{i=1}^{9}(\mathbf{o}_{i}-\mathbf{g}_{i})^2,
\label{eq:loss}
\end{equation}
where $i$ represents the indices of the $3 \times 3$ particle patches in the target cloth and $N$ is the total number of inferences made by the DNN model. The DNN model was trained by using the adaptive moment estimation optimization method~\cite{kingma2014adam} until the loss plateaued. We set the learning rate to 0.0001 and batch size to 50,000. 




Table~\ref{tab:training_data} shows the amount of training data that comprised the input features and ground-truth output pairs collected to train the upscaling DNN in each example and the corresponding number of frames. As shown in the table, training data was collected for more than four million pairs in all three examples. To collect the training data, we carried out target cloth simulations with P-ADMM (projective dynamics applying the ADMM)~\cite{7990052}. The training data contain results from various simulations where the mass value, the spring constant, and the time step were varied. The upscaling DNN was trained separately for each example, and the training time was approximately eight hours, respectively.


\section{Target Cloth Simulation Generation}
\label{sec:proposed-method}


Our cloth simulation system uses the results of miniature cloth simulation and an upscaling DNN to simulate the target cloth efficiently. An overview of our system is presented in Figure~\ref{fig:overview}. First, the miniature cloth system is created by down-sampling and down-scaling operations on the target cloth system (as in the data generation process). Then, physically-based simulations are performed on the miniature cloth system. Finally, the results of target cloth simulation are generated by the upscaling DNN inference, which takes the results of miniature cloth simulations as input feature vectors. In this process, because the miniature cloth system is a down-scaled version of the target cloth system, the external forces in the miniature cloth system should also be down-scaled.

\subsection{Relationship between Miniature and Target Cloth Simulation}
\label{subsec:relationship}

Consider a mass-spring system with positions $\mathbf{x}$, velocities $\mathbf{v}$, a mass-matrix $\mathbf{M}$, and time step $h$. The next positions of the target cloth $\mathbf{x}_{t+1}$ are calculated by using the following implicit integration equations:
\begin{equation}\label{eq:implicit_integration}
\mathbf{x}_{t+1}-(\mathbf{x}_t+h\mathbf{v}_t)=\mathbf{M}^{-1} h^2 \left ( f_{I}( \mathbf{x}_{t+1})+f_{E}\right ),
\end{equation}
where $f_{I}$ represents the internal forces of the cloth model and $f_{E}$ represents the external forces~\cite{Liu:2013:FSM:2508363.2508406}.
%
%
Because the miniature cloth system is a version of the target cloth system down-scaled by a factor of two,
$\mathbf{x}_t=2\mathbf{x}'_t$ and $h\mathbf{v}_t=\mathbf{x}_t-\mathbf{x}_{t-1}=2h\mathbf{v}'_t$, where $\mathbf{x}'_t$ and $\mathbf{v}'_t$ are the current positions and the velocities in the miniature cloth system, respectively.
Based on (\ref{eq:implicit_integration}), we can express the update rule for the miniature cloth system as follows:
\begin{equation}\label{eq:implicit_integration2}
2\mathbf{x}'_{t+1}-2(\mathbf{x}'_t+h\mathbf{v}'_t)=\mathbf{M}^{-1} h^2 \left ( f_{I}(2{\mathbf{x}'_{t+1}})+f_{E}\right ).
\end{equation}


\begin{figure*}[htbp]
\centering
\begin{subfigure}{0.48\textwidth}
\includegraphics[width=\columnwidth]{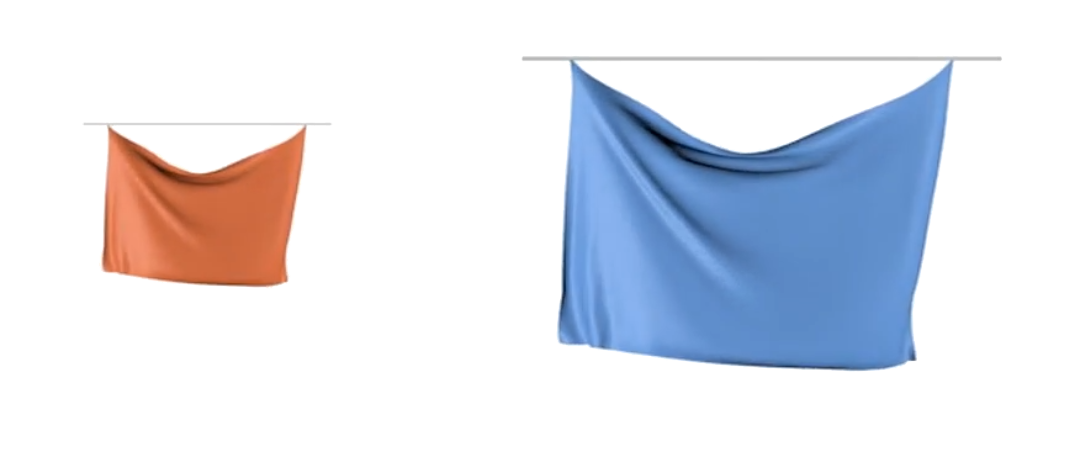}
\caption{Curtain}
\end{subfigure}
\begin{subfigure}{0.48\textwidth}
\includegraphics[width=\columnwidth]{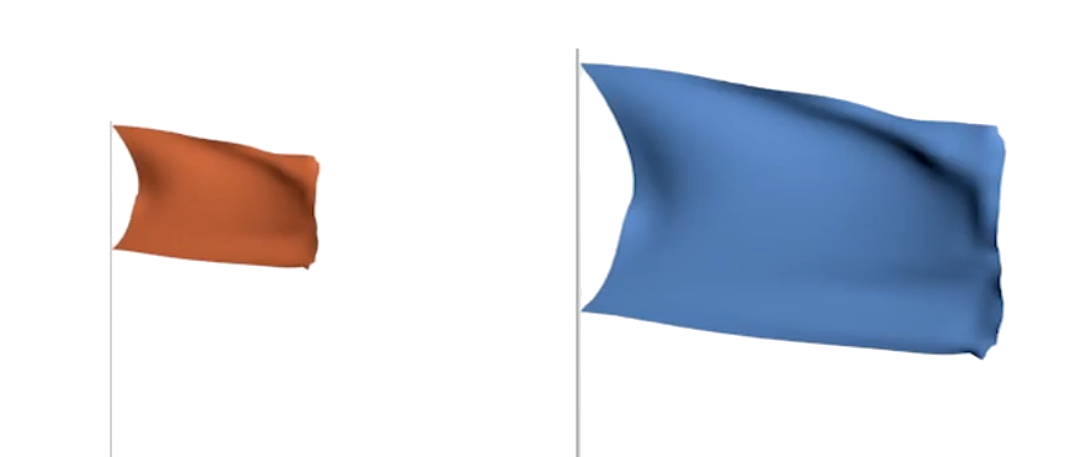}
\caption{Flag}
\end{subfigure}
\begin{subfigure}{0.48\textwidth}
\includegraphics[width=\columnwidth]{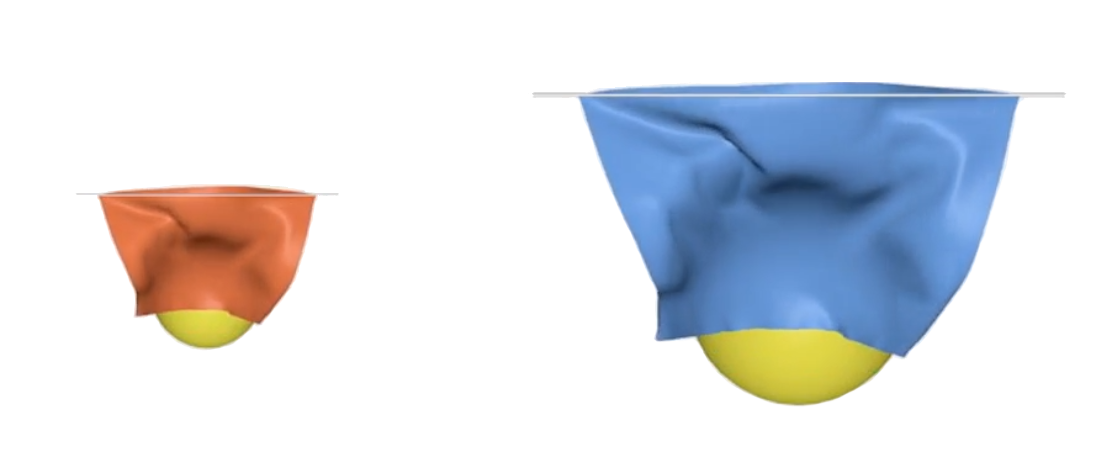}
\caption{Collision}
\end{subfigure}
\caption{Results of target cloth simulation achieved by upscaling DNN inference based on miniature cloth simulation: (a) curtain, (b) flag, and (c) collision. (Left) Miniature cloth. (Right) Target cloth.}
\label{fig:upscaling}
\end{figure*}

\noindent
Assuming that each spring follows Hooke's law~\cite{halliday2013}, the internal force can be expressed as
\begin{equation}
f_{I}(2\mathbf{x}'_{t+1}) = 2k\Delta L = 2f_{I}(\mathbf{x}'_{t+1}),
\label{eq:spring_internal}
\end{equation}
where $k$ is the spring constant and $\Delta L$ is the spring displacement. By substituting (\ref{eq:spring_internal}) into (\ref{eq:implicit_integration2}), we can derive the final updated rule for miniature cloth simulation as follows:
\begin{equation}\label{eq:implicit_integration_miniature}
\mathbf{x}'_{t+1}-(\mathbf{x}'_t+h\mathbf{v}'_t)=\mathbf{M}^{-1} h^2 \left ( f_{I}(\mathbf{x}'_{t+1})+\frac{1}{2}f_{E}\right ),
\end{equation}
which indicates that the external force of the miniature cloth system should be set to $1/2$ of that of the target cloth system. By down-scaling the external forces, the movements of the miniature cloth obtained by simulation are similar to those obtained by target cloth simulation.

\subsection{Inference of Upscaling Deep Neural Networks}
\label{subsec:upscaing-dnn-inference}

Upscaling DNN infers the positions of the particles at frame $t$ of the target cloth using the positions of the particles at frames $t$, $t-1$, and $t-2$ of the miniature cloth simulation as input feature vectors. In this process, the inference of the DNN is performed in patch units, and, thus, it is possible to accelerate this computation through parallel processing based on OpenMP~\cite{Dagum:1998:OIA:615255.615542}. 

\begin{figure}[htbp]
  \centering
  \includegraphics[width=\columnwidth]{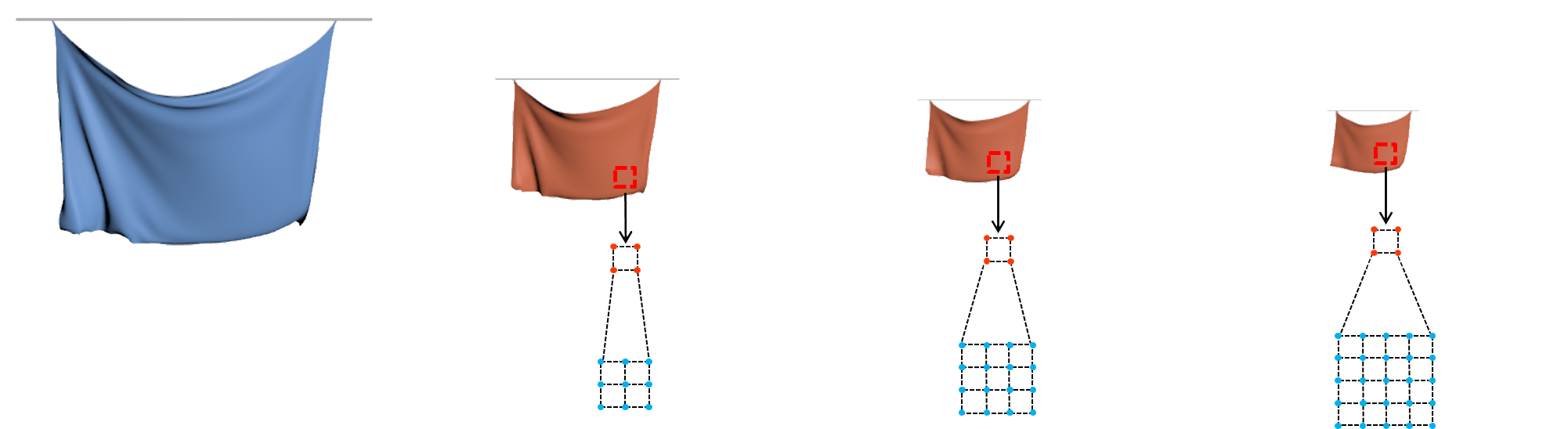}
  \caption{Miniature cloth created with different DSDS factors. From left to right: target cloth and DSDS factors 2, 3, and 4.}
  \label{fig:factor_miniature}
\end{figure}

\begin{figure}[htbp]
  \centering
  \includegraphics[width=\columnwidth]{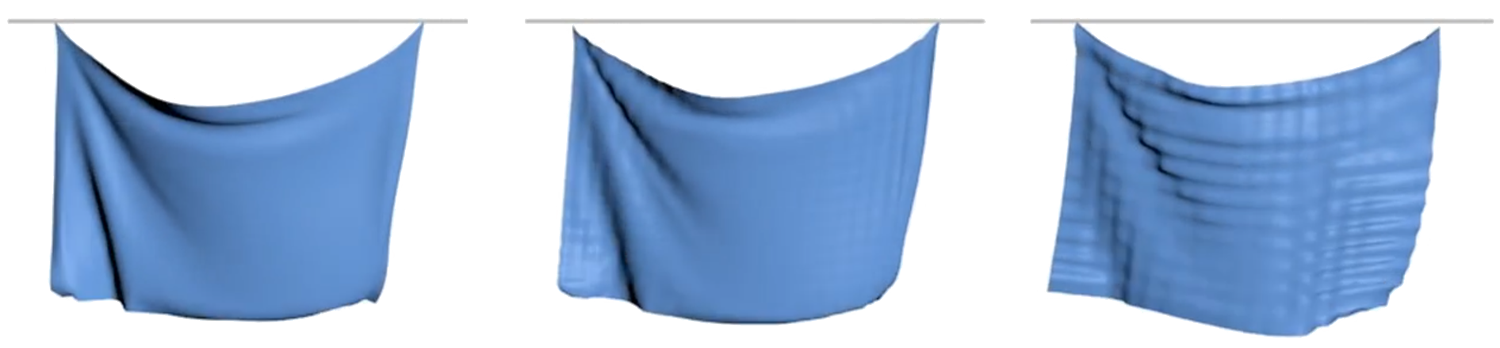}
  \caption{Target cloth generated with different DSDS factors. From left to right: DSDS factors 2, 3, and 4.}
  \label{fig:factor}
\end{figure}

\section{Results}
\label{sec:results}


Our DNN model was configured and trained using Tensorflow~\cite{tensorflow2015-whitepaper} on a PC with a quad-core Intel Core i7-3770 CPU with eight threads, 32 GB of RAM, and an NVIDIA GTX 970 GPU. We used GPU acceleration for the training process. However, because the P-ADMM and H-DNN methods used to compare performance do not use GPU acceleration, we did not use GPU acceleration in the testing of our approach; thus, we made the comparison under the same conditions.
In all three methods, parallel processes based on OpenMP~\cite{Dagum:1998:OIA:615255.615542} were used for the test processes.
In addition, we constructed the inference machine by moving the weights and bias values of the neural network model trained in Python-based Tensorflow to the C++-based deep learning framework tiny-dnn~\cite{Tinydnn}. This shortened the inference time and allowed simulations in conjunction with the P-ADMM library~\cite{7990052}, which is written in C++.


Figure~\ref{fig:upscaling} shows the results of miniature cloth simulation calculated using P-ADMM and the results of target cloth simulation generated by the upscaling DNN inference. The red ones are the miniature cloth, and the blue ones are the target cloth. The proposed method generated stable target cloth simulations under various conditions in all three examples (curtain, flag, and collision).

\subsection{Performance According to Features}
\label{subsec:performance-features}

We compared the performance of the three different cases by changing only the input features of the proposed DNN structure to identify the appropriate input feature combination for the upscaling DNN. Three DNN models were trained with three types of feature combinations: the location of the particles in the current frame of a miniature patch, the location and velocity of the particles in the current frame, and the locations of the particles in three consecutive frames.  
The training and test data for the three cases were prepared with 834,861 input data and 122,388 output data, respectively, in identical simulations of the flag example.
Table~\ref{tab:feature_comparison} shows the training and test errors for three different input combinations. The error is the value of the loss function in Equation (\ref{eq:loss}). 
Training is stable in all three cases, but, when using three consecutive frames of data as an input feature, the test error is lower than the other two cases.

\begin{table}[t]
\centering
\caption{Training and test MSE according to the input features of the upscaling DNN: the positions of the particles in the current frame, the positions and velocities of the particles in the current frame, and the positions of the particles in three consecutive frames (from top to bottom).}
\label{tab:feature_comparison}
\renewcommand{\arraystretch}{1.2}
\begin{tabular}{ccc}
\hline
Features             & Training error & Test error \\
\hline
Positions            & 9.094e-05       & 2.85e-04   \\
Positions + Velocities & 8.876e-05       & 2.778e-04   \\
Positions in 3 frames       & 8.856e-05       & 2.691e-04   \\     
\hline
\end{tabular}
\end{table}

\begin{figure*}[htbp]
\centering

\begin{subfigure}{0.24\textwidth}
\includegraphics[width=\columnwidth]{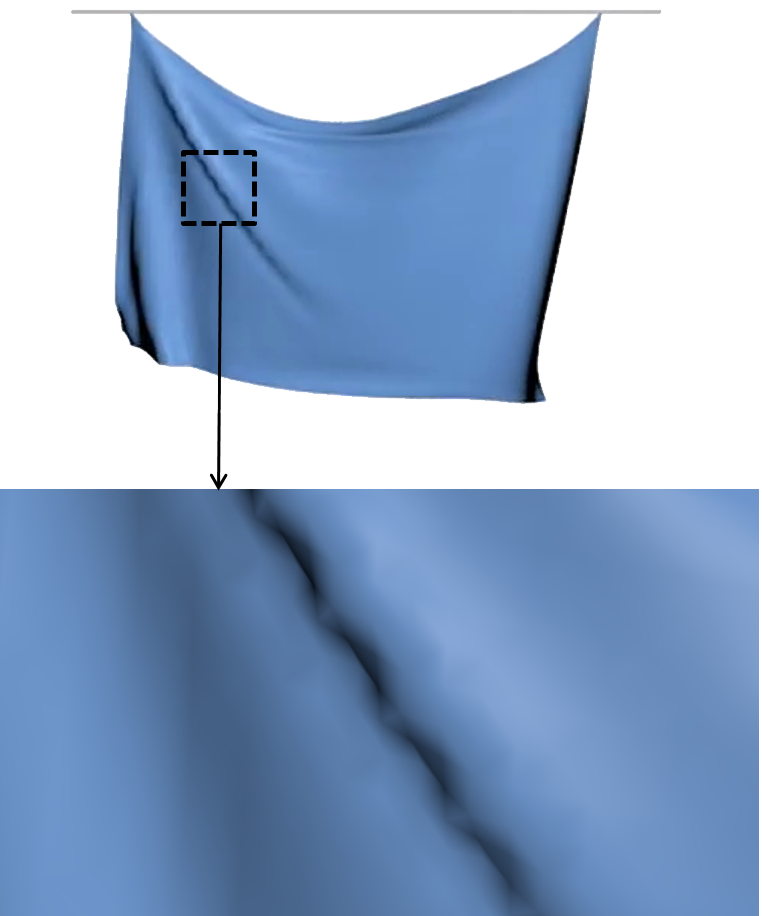}
\caption{Bilinear}
\end{subfigure}
\begin{subfigure}{0.24\textwidth}
\includegraphics[width=\columnwidth]{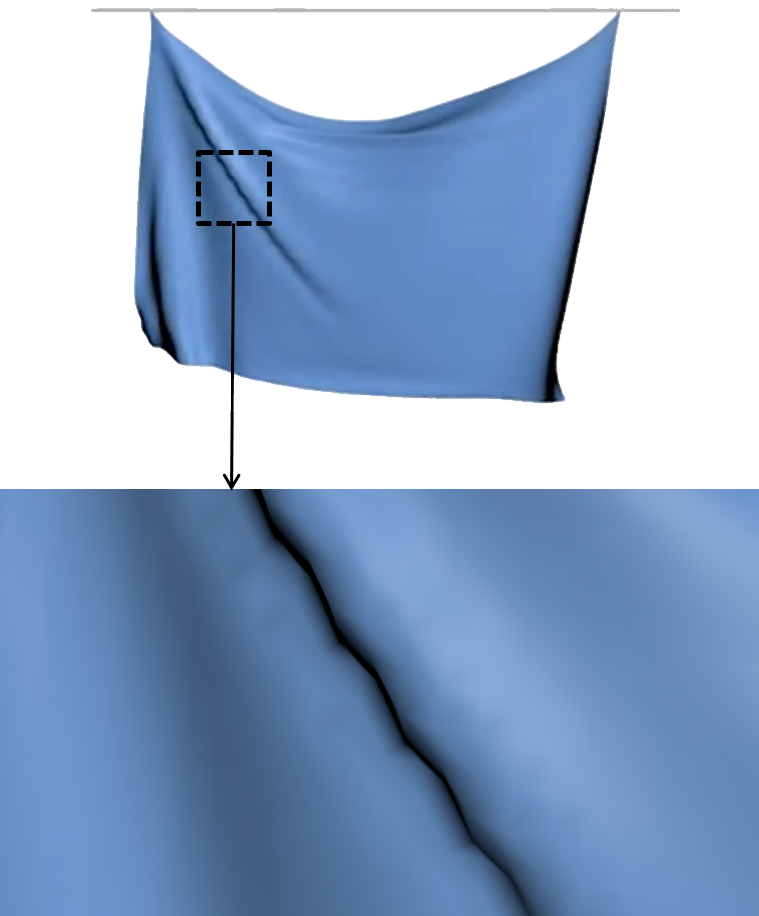}
\caption{Biquadratic}
\end{subfigure}
\begin{subfigure}{0.24\textwidth}
\includegraphics[width=\columnwidth]{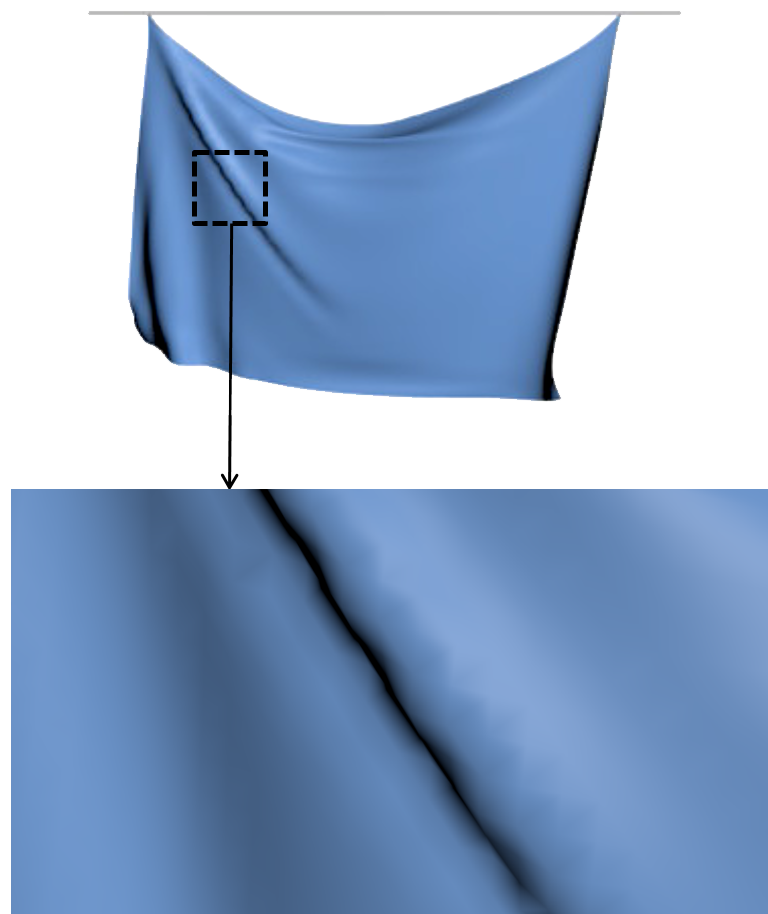}
\caption{Bicubic}
\end{subfigure}
\begin{subfigure}{0.24\textwidth}
\includegraphics[width=\columnwidth]{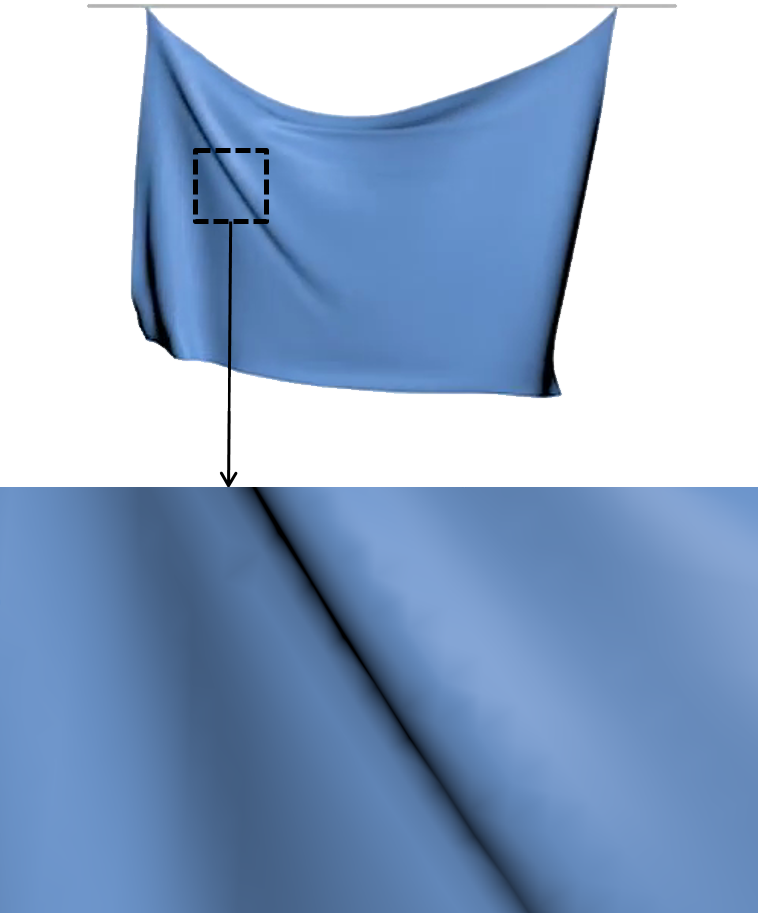}
\caption{Upscaling DNN}
\end{subfigure}

\caption{Generated results according to upscaling methods: (a) bilinear interpolation, (b) biquadratic interpolation, (c) bicubic interpolation, and (d) upscaling DNN.}
\label{fig:upsampling}
\end{figure*}

\subsection{Performance According to DSDS Factor}

We investigated how the performance of the proposed method was affected by increasing DSDS factor (2, 3, and 4) used to construct the miniature model from the original model (see Figure~\ref{fig:factor_miniature}). Considering the relationship between the miniature and target cloth systems shown in Section~\ref{subsec:relationship}, the external forces in the miniature cloth system should also be down-scaled. That is, with DSDS factors 3 and 4, external forces were set to 1/3 and 1/4 of the external forces given in the target system, respectively. For DSDS 3 and 4, more than four million pairs of data were collected, and 80\% and 20\% of the data were used in training and testing, respectively, in the case of DSDS factor 2. As shown in Table~\ref{tab:factor}, the larger the factor, the better the time performance because of the decreasing inference time, but the per particle Euclidean distance error between the inferred position and the ground-truth output position increased. The results generated by the inference of the trained upscaling DNN under each DSDS factor are shown in Figure~\ref{fig:factor}. As shown in the figure, as the DSDS factor increased, the quality of the results decreased.

\begin{table}[t]
\centering
\caption{Euclidean distance error per particle and computation time per frame (msec) according to different DSDS factors.}
\label{tab:factor}
\renewcommand{\arraystretch}{1.2}
\begin{tabular}{ccc}
\hline
DSDS factor & Euclidean error & Time per frame (ms) \\
\hline
2      & 1.816e-03   & 37.5               \\
3      & 2.916e-03   & 19.1               \\
4      & 3.901e-03   & 16.2               \\
\hline
\end{tabular}
\end{table}

\subsection{Performance According to Network Structures}


We compared the performance of various DNN models to determine the most suitable structure of the upscaling DNN in the proposed method.
Using the dataset constructed in Section~\ref{sec:upscalingDNN}, we used 80\% as training data and 20\% as test data in each example, respectively. 
Table~\ref{tab:model_comparison} shows the test error of various DNN models, where U*FC-V means the DNN has U hidden layers with V nodes at each layer; for example, the 4*FC-64 model is the DNN structure with four hidden layers and 64 nodes at each layer that is proposed in Section~\ref{subsec:DNN_structure}. Compared to the 5*FC-64 model with an increased number of layers and the 3*FC-64 model with a decreased number of layers, 4*FC-64 showed the lowest test error when trained with the same training data. In addition, when compared with the 4*FC-128 and 4*FC-32 models with increased and decreased, respectively, numbers of nodes per layer, 4*FC-64 also showed the lowest test error.

\begin{table}[t]
\centering
\caption{Test error of DNN models with different structures. U*FC-V means the DNN has U hidden layers with V nodes at each layer.}
\label{tab:model_comparison}
\renewcommand{\arraystretch}{1.2}
\begin{tabular}{cccc}
\hline
DNN model & Curtain   & Flag      & Collision \\
\hline
4*FC-32   & 9.594e-06 & 1.674e-04 & 1.66e-05  \\
3*FC-64   & 6.405e-06 & 1.276e-04 & 1.357e-05 \\
4*FC-64   & 6.214e-06 & 1.217e-04 & 1.17e-05  \\
5*FC-64   & 7.158e-06 & 1.677e-04 & 1.454e-05 \\
4*FC-128  & 6.772e-06 & 1.486e-04 & 1.414e-05 \\
\hline
\end{tabular}
\end{table}

\subsection{Comparison with Interpolation Methods}

We also compared the performance when using conventional interpolation methods and DNN for the upscaling process. Bilinear, biquadratic, and bicubic methods~\cite{press2007numerical} were used for the interpolation. As shown in Table~\ref{tab:upscaling_error}, the use of interpolation methods improved the computation time performance but increased the Euclidean distance error slightly. Furthermore, the interpolation methods generated undesirable effects when expressing detailed wrinkles, as can be seen in the large wrinkles starting from the upper left part of the cloth (see Figure~\ref{fig:upsampling}). When using interpolation-based methods, it is difficult to maintain a topology that considers the relationship between particles in the wrinkle portion where self-intersection occurs. However, the DNN model can produce similar results to P-ADMM because training maintains a relatively good topology between particles in the wrinkle portion.

\begin{table}[t]
\centering
\caption{Euclidean distance error (per particle) and computation time per frame (msec) according to the upscaling method.}
\label{tab:upscaling_error}
\renewcommand{\arraystretch}{1.2}
\begin{tabular}{ccc}
\hline
Upscaling   & Euclidean error & Time per frame (ms) \\
\hline
Bilinear    & 1.826e-03           & 17.7                \\
Biquadratic & 1.825e-03           & 17.7                \\
Bicubic     & 1.825e-03           & 17.8                \\
DNN         & 1.816e-03           & 37.5                \\
\hline
\end{tabular}
\end{table}

\subsection{Comparisons with Conventional Methods}

\begin{figure}[tbp]
  \centering
  \includegraphics[width=\columnwidth]{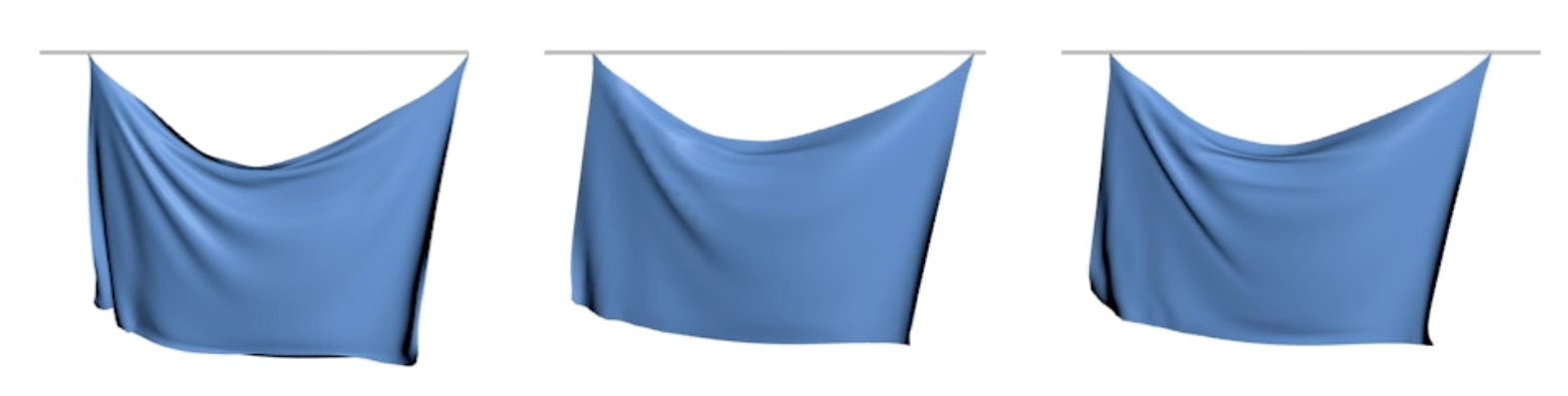}
  \caption{Results of simulation of the curtain example. From left to right: P-ADMM, H-DNN, and the proposed method.}
  \label{fig:curtain}
\end{figure}

\begin{figure}[tbp]
  \centering
  \includegraphics[width=\columnwidth]{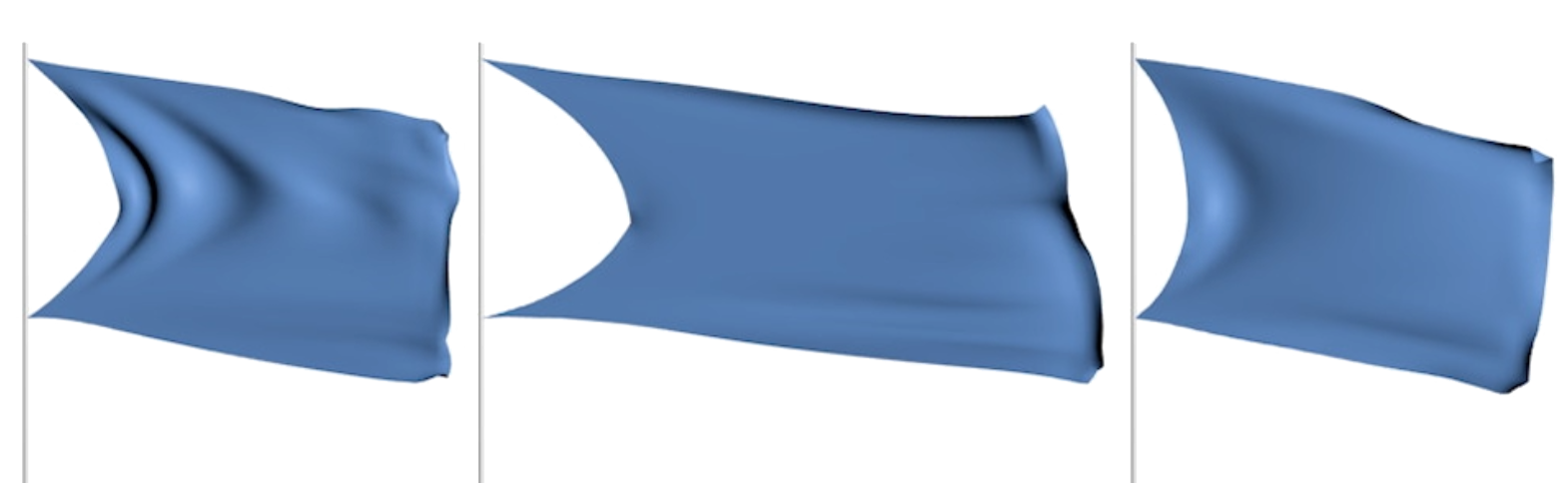}
  \caption{Results of simulation of the flag example. From left to right: P-ADMM, H-DNN, and the proposed method.}
  \label{fig:flag}
\end{figure}

\begin{figure}[tbp]
  \centering
  \includegraphics[width=\columnwidth]{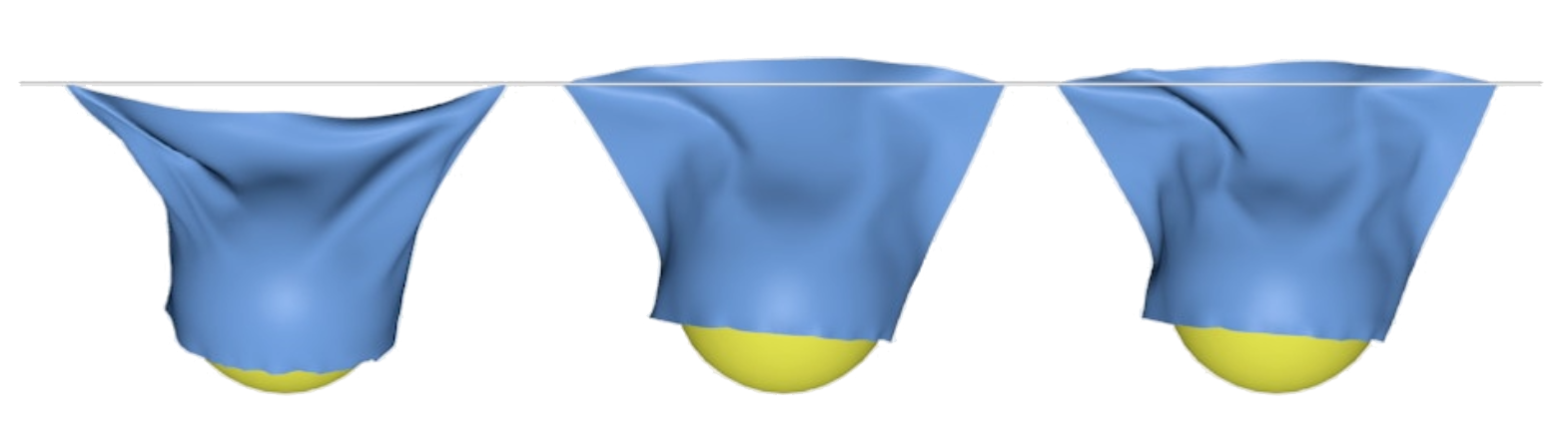}
  \caption{Results of simulation of the collision example. From left to right: P-ADMM, H-DNN, and the proposed method.}
  \label{fig:collision}
\end{figure}

We compared the results obtained using our proposed method with those from the existing method for verification. The results of the simulation of each example using P-ADMM, H-DNN~\cite{Oh:2018:HCS:3208159.3208162}, and the proposed method are shown in Figure~\ref{fig:curtain}, \ref{fig:flag}, \ref{fig:collision}. All three methods produced stable simulations. However, the results of the proposed method are more similar to those of P-ADMM than those obtained using H-DNN. As shown in Figure~\ref{fig:curtain}, the proposed method is more similar to P-ADMM than H-DNN in the upper left part of the generated cloth. In particular, the proposed method preserves the high-frequency details of P-ADMM better than H-DNN in the collision parts of the cloth (see also Figure~\ref{fig:collision}). In particular, in the flag example (see Figure~\ref{fig:flag}) with applied wind force, the use of H-DNN resulted in a significantly different movement from that obtained using P-ADMM. This is because the wind force is modeled as a function using the triangle area, the normal, and the tangent vectors of the cloth model~\cite{Wejchert:1991:AA:122718.122719}. H-DNN does not preserve the triangle area of the target cloth, so the wind force is computed incorrectly. However, because the proposed method preserves the size of the target cloth's triangles, the results are more similar to those generated by the P-ADMM, and the detailed high-frequency movement is better represented.

\begin{figure}[tbp]
\centering

\begin{subfigure}{0.48\textwidth}
\includegraphics[width=\columnwidth]{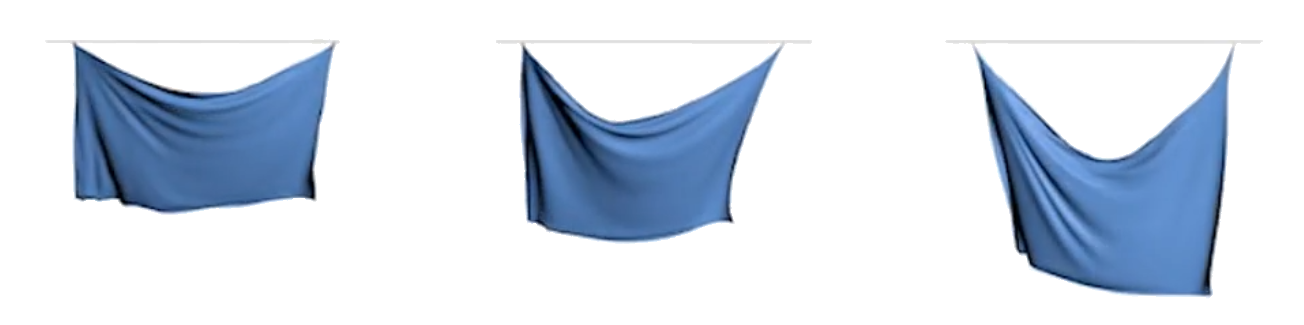}
\caption{P-ADMM}
\end{subfigure}
\begin{subfigure}{0.48\textwidth}
\includegraphics[width=\columnwidth]{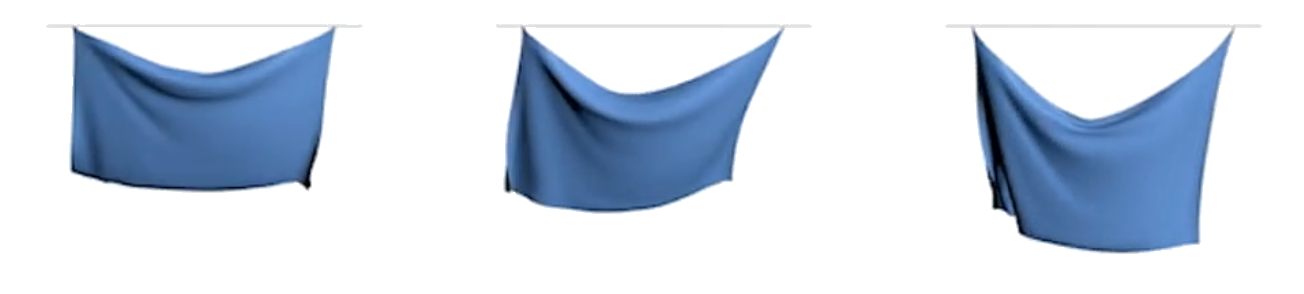}
\caption{H-DNN}
\end{subfigure}
\begin{subfigure}{0.48\textwidth}
\includegraphics[width=\columnwidth]{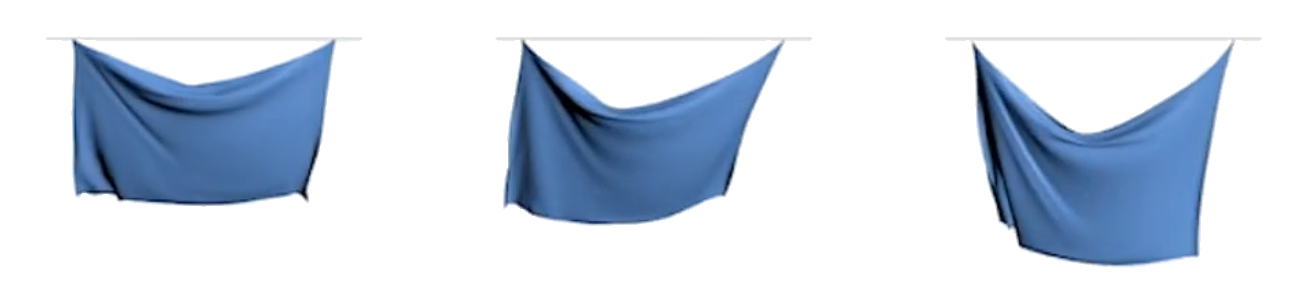}
\caption{Proposed Method}
\end{subfigure}

\caption{Results of simulation when the mass value is changed: (a) P-ADMM, (b) H-DNN, and (c) the proposed method. Mass values from left to right: 1.5, 3.0, 4.5.}
\label{fig:mass}
\end{figure}

\begin{figure}[tbp]
\centering

\begin{subfigure}{0.48\textwidth}
\includegraphics[width=\columnwidth]{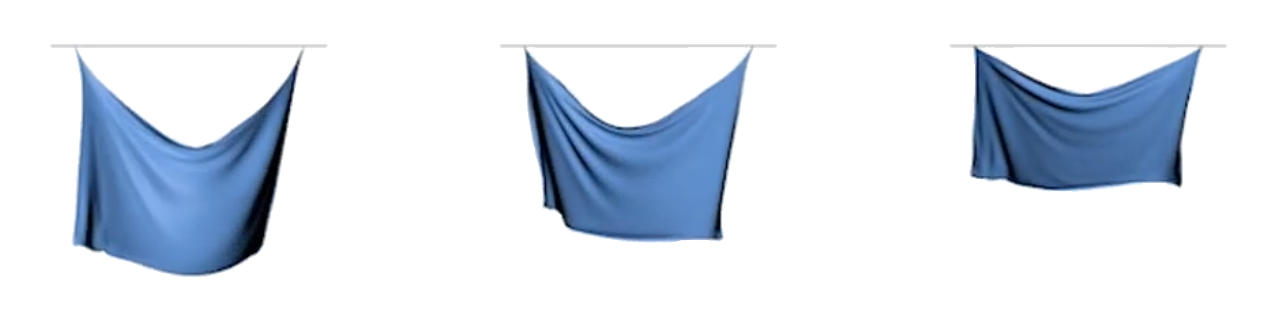}
\caption{P-ADMM}
\end{subfigure}
\begin{subfigure}{0.48\textwidth}
\includegraphics[width=\columnwidth]{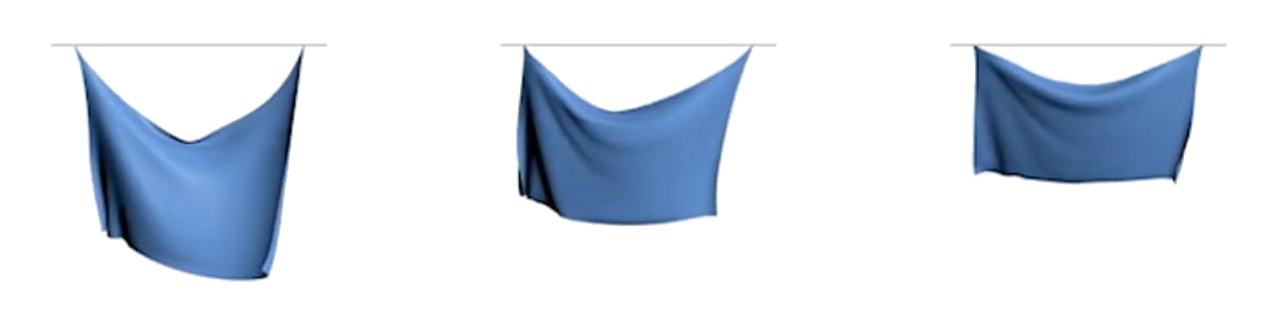}
\caption{H-DNN}
\end{subfigure}
\begin{subfigure}{0.48\textwidth}
\includegraphics[width=\columnwidth]{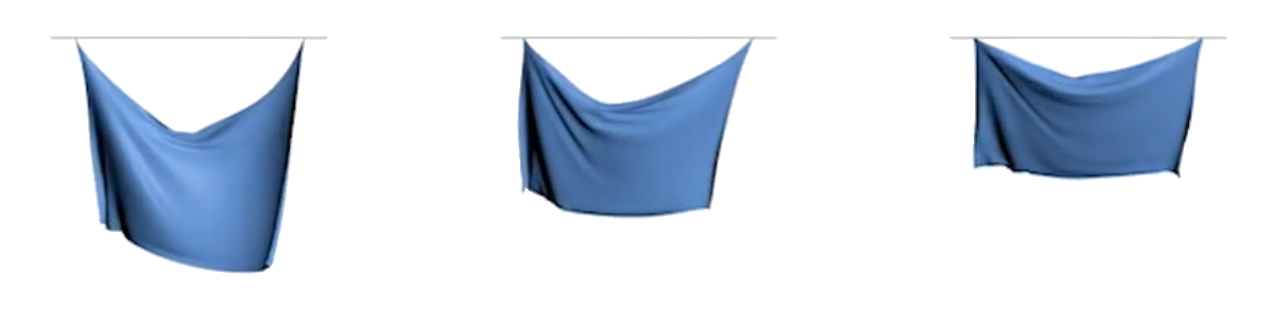}
\caption{Proposed Method}
\end{subfigure}

\caption{Results of simulation when the spring constant is changed: (a) P-ADMM, (b) H-DNN, and (c) the proposed method. Spring constants from left to right: 40, 70, 150.}
\label{fig:spring}
\end{figure}

The proposed method even produces stable results when the physical properties of the cloth are altered, such as the mass of a particle and the spring constant. Figure~\ref{fig:mass} and Figure~\ref{fig:spring} show the results of simulation using P-ADMM, H-DNN, and the proposed method when the mass and spring constant are altered, respectively. The proposed method always generates stable results, even when the mass and spring constant are changed. The proposed method also preserves more high-frequency details compared to H-DNN (see Figure~\ref{fig:mass} and Figure~\ref{fig:spring}).

We compared the per particle Euclidean distance error to verify that the results of the proposed method are more similar to the results of P-ADMM compared to the results of H-DNN. Table~\ref{tab:accuracy_comparison} shows the per particle Euclidean distance error with P-ADMM result in each example. Because the proposed method generates results with much better preservation of the high-frequency details, in all examples, it has a lower error than H-DNN. In particular, in the flag example, the proposed method shows much less error than H-DNN because the results of the proposed method are stretched to a similar extent as the results of the P-ADMM.

\begin{table}[t]
\centering
\caption{Per particle Euclidean distance error with P-ADMM result.}
\label{tab:accuracy_comparison}
\renewcommand{\arraystretch}{1.2}
\begin{tabular}{cccc}
\hline
Example   & H-DNN & Proposed method \\
\hline
Curtain   & 1.841e-03 & 1.816e-03           \\
Flag      & 2.492e-02 & 7.767e-03           \\
Collision & 5.982e-03 & 5.973e-03           \\    
\hline
\end{tabular}
\end{table}

\begin{table*}[t]
\centering
\caption{Comparison of computation time per frame (msec) for P-ADMM, H-DNN, and the proposed method. The curtain and flag examples were rendered with a time step of 40 msec and the collision example was rendered with a time step of 20 msec.}
\label{tab:time_comparison}
\renewcommand{\arraystretch}{1.2}
\begin{tabular}{ccccc}
\hline
Example & \#Particles & \begin{tabular}[c]{@{}c@{}}P-ADMM\\(ms)\end{tabular} & \begin{tabular}[c]{@{}c@{}}H-DNN\\(ms)\end{tabular} & \begin{tabular}[c]{@{}c@{}}Proposed method\\(ms)\end{tabular} \\ \hline
Curtain    & 3577   & 72.8                                                  & 49.7                                                & 37.5                                                      \\ 
Flag   & 1617   & 33.2                                                  & 24.8 & 18.1   \\ 
Collision & 2501 & 50.2                                                  & 35.1 & 33.2   \\ 
\hline
\end{tabular}
\end{table*}


We compared the time performance to verify that the proposed method generates results efficiently. Table~\ref{tab:time_comparison} lists the mean times required to generate one frame of simulation for P-ADMM, H-DNN, and the proposed method. For the simulation of a cloth containing the same number of particles, the proposed method generates results more efficiently than the other two methods. This is because the proposed method uses a small number of particles and a lightweight network model. Additionally, the inference process of the proposed method is performed in patch units, as compared to H-DNN's triangle units. Therefore, the number of inferences is smaller, resulting in more efficient computation time.

\section{Conclusion}
\label{sec:conclusion}


We have proposed an efficient cloth simulation method using miniature cloth simulation and upscaling DNN. The proposed method performs physically-based simulation in a miniature cloth system created by applying down-sampling and down-scaling operations to the target cloth. Using the results, the upscaling DNN generates the target cloth simulation. Because this method uses a miniature cloth with a small number of particles and a lightweight network model, efficient target cloth simulation is possible. In addition, the miniature cloth retains the physical properties of the target cloth. In other words, high-frequency details are preserved better than when using H-DNN because the results are generated using miniature cloth simulations that move in a similar way to the target cloth simulation. We have demonstrated experimentally that the proposed method can generate fast and stable simulations under various conditions.

\begin{figure}[tbp]
  \centering
  \includegraphics[width=\columnwidth]{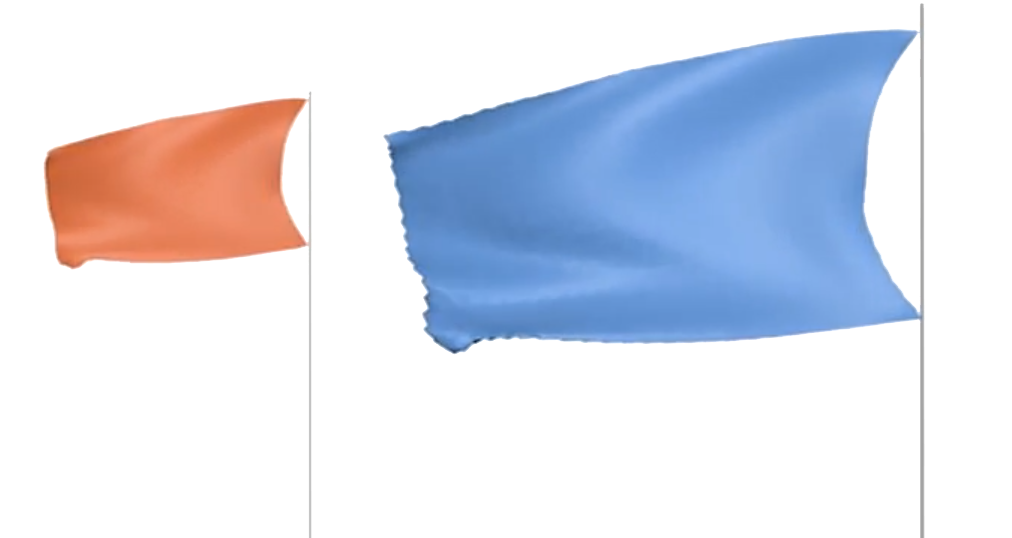}
  \caption{Failure case for the proposed method. These are the results of a simulation with a wind direction opposite to the wind direction in the training data.}
  \label{fig:limitation}
\end{figure}

Figure~\ref{fig:limitation} presents the results generated by the proposed method when simulating a wind direction that is opposite to the wind direction in the training data. As shown in the figure, unwanted wrinkles are formed on the cloth as a result of incorrect inference by the DNN. The proposed method has a limitation in that it is difficult to make accurate inferences about situations that do not exist in the training data. This is a common limitation of learning-based methods that is typically solved by using more training data. Therefore, it is essential to construct training data that is suitable for a wide variety of situations. We plan to generalize the proposed method to other physically-based simulation fields, such as fluid or soft-body simulation. Recently, novel network structures with higher accuracy compared to plain networks have been introduced in the image classification field~\cite{7780459, 8099726}. We are considering using these novel network structures to construct a better model that reduces the differences from the results of ground-truth simulation.

\bibliographystyle{ACM-Reference-Format}
\bibliography{sample-bibliography}

\end{CJK}

\end{document}